%% file: recsys.tex
\documentclass{llncs}
\usepackage{paralist}
\usepackage{amsmath}
\usepackage{graphicx}
\usepackage{subfigure}
\usepackage{xspace}
\newcommand{\norm}[1]{\|#1\|}
\newcommand{\pair}[1]{\langle#1\rangle}

\newcommand{\pathto}[1]{\mathop{\rightarrow}\limits^{#1}}
\newcommand{\mat}[1]{\textbf{\emph{#1}}}

\newcommand{\derivation}[2]{\frac{\partial #1}{\partial #2}}

\newcommand{\diag}[1]{\textsf{diag}(#1)}
\newcommand{\pathsim}{PathSim\xspace}

\newcommand{\myfootnote}[1]{\footnote{\small #1}}
\newcommand{\myparagraph}[1]{\vspace{1ex}\noindent\textbf{#1.}\hspace{1em}}

\begin{document}
\title{Content-Based Top-$N$ Recommendation using Heterogeneous Relations}
\author{Yifan Chen\inst{1} \and Xiang Zhao\inst{1} \and Junjiao Gan\inst{2} \and Junkai Ren\inst{1} \and Yang Fang\inst{1}}
\institute{National University of Defense Technology, China
\and Massachusetts Institute of Technology, United States}

\thispagestyle{empty}
\maketitle
\begin{abstract}
Top-$N$ recommender systems have been extensively studied. However, the sparsity of user-item activities has not been well resolved. While many hybrid systems were proposed to address the \emph{cold-start} problem, the profile information has not been sufficiently leveraged. Furthermore, the heterogeneity of profiles between users and items intensifies the challenge. In this paper, we propose a content-based top-$N$ recommender system by learning the global term weights in profiles. To achieve this, we bring in \pathsim, which could well measures the node similarity with heterogeneous relations (between users and items). Starting from the original TF-IDF value, the global term weights gradually converge, and eventually reflect both profile and activity information. To facilitate training, the derivative is reformulated into matrix form, which could easily be paralleled. We conduct extensive experiments, which demonstrate the superiority of the proposed method.
\end{abstract}

\input{intro}

\input{relat}

\input{sim}
\input{learn}
\input{exp}

\section{Conclusion} \label{sec:conclude}
In this paper, we proposed a content-based top-$N$ recommender system by leveraging item profiles. 
We first employed \pathsim to measure the item similarity on the top of heterogeneous relations between users and items, and then optimized the global term weights towards the \pathsim similarities.
To facilitate training, the derivation was reformulated into matrix form, which could easily be paralleled. We conducted extensive experiments, and the experimental results demonstrate the superiority of the proposed method.

\small
\bibliographystyle{splncs03}
\bibliography{recsys}

\end{document}

%% file: intro.tex
\setcounter{footnote}{0}
\section{Introduction}

Recommender systems typically leverage two types of signals to effectively recommend \emph{items} to \emph{users} - user activities, and content matching between user and item profiles. Depending on what to use, the recommendation models in literature are usually categorized into collaborative filtering, content-based and hybrid models~\cite{balabanovic1997fab}.
In real-world applications, solely employing collaborative filtering or content-based models can not achieve desirable results, as is often the case that single source of information tends to be incomplete.

To better illustrate, we motivate the following example in architecture.
Recently, \texttt{Vanke}, a leading real-estate corporation in China, started a Uber-For-Architects project, namely \texttt{NOSPPP}, which tries to match architects with appropriate projects based on previous project information of architects and firms. During the running of \texttt{NOSPPP}, the data collected are two-fold, with the participated projects and the resumes, respectively.
In terms of recommender system, the former is named ``feedback'' (of users) while the latter is referred as ``profiles'' (of items).
Due to the sparsity of feedback, collaborative filtering based recommenders would face the \emph{cold-start} problem, and hence, we have to resort to content-based or hybrid models~\cite{DBLP:conf/www/GuZHS16}. However, unlike the applicant-job scenario therein, where the profiles of users and jobs could well match, the profiles of architects and projects describe things in two different worlds. Specifically, the profiles of architects presents the working experience and skills, while the profiles of projects tells the area, the interior and exterior constructions, etc. This is rational, as designing architectures is in the form of art, where it is hard to specify the conditions or requirements through decomposition.
In response to applications like \texttt{NOSPPP}, we explore recommendation utilizing both \emph{sparse feedback} and \emph{heterogeneous profiles}.

In this paper, we exploit a hybrid recommendation method that ensembles both sources of information.
For ease of exposition, we consider the case that auxiliary information exists only on the side of items, and propose a item-based top-$N$ recommendation algorithm~\myfootnote{Without loss of generality, it is straightforward to extend the idea to the case of auxiliary information on both sides of users and items.}.
Classic item-based collaborative filtering uses the direct link information for recommendation without diffusing the influence of other user-item links.
By regarding user-item interactions as a bi-type information network, 
%
we observe that such influence can be captured by item node similarity, where \pathsim~\cite{DBLP:journals/pvldb/SunHYYW11} via \emph{meta-path} is served.
%
Moreover, while content matching between heterogeneous profiles of users and items does not produce explicable results, methods including~\cite{wang2015unsupervised,DBLP:journals/tip/WangLWZZH15,DBLP:conf/mm/WangLWZ15,DBLP:conf/sigir/WangLWZZ15,DBLP:conf/mm/WangLWZZ14} suggest high similarity among objects within the same subspace, thus we contend that it can be employed for matching profiles between item-item or user-user, since profiles of same type is naturally homogeneous.
A standard way to measure similarity between two profiles is computing the cosine similarity of the two bags of words, and each word is weighted by term frequency $tf$ (within the document) $\times$ inverted document frequency $idf$ (of the term within the corpus).
While the local term frequency could be computed offline, it has been suggested~\cite{DBLP:conf/www/GuZHS16} that the global term weights $idf$ requires further optimization to achieve better precision. Thus, we optimize the global term weights with the guidance of the similarity from \pathsim. 

In summary, the major contribution of the paper is a novel hybrid recommendation method, the overview of which is outlined as follows:
\begin{enumerate}[(1)]
  \item Derive item similarity measured by meta-paths using \pathsim;
  \item Optimize the global term weights guided by \pathsim; and
  \item Recommend top-$N$ items based on nearest neighbor collaborative filtering.
\end{enumerate}

\myparagraph{Organization}
Section~\ref{sec:related} discusses related work. We present the method for deriving initial similarity between items in Section~\ref{sec:network}, and then, introduce the learning method for optimizing global term weights in Section~\ref{sec:learn}. Experiment results are in Section~\ref{sec:exp}, followed by conclusion in Section~\ref{sec:conclude}.

%% file: relat.tex
\section{Related Work} \label{sec:related}

Top-$N$ recommender systems have been extensively studied during the last few years, which could be classified into three categories.

The first category is neighborhood-based collaborative filtering, which could be further classified into three classes: item-based and user-based. Given a certain user, \emph{user-based-nearest-neighbor} (\textsf{userkNN})~\cite{DBLP:conf/cikm/Karypis01,DBLP:journals/tois/DeshpandeK04,papagelis2005qualitative} first identifies a set of similar users, and then recommends top-$N$ items based on what items those similar users have purchased. Similarly, \emph{item-based-nearest-neighbor} (\textsf{itemkNN})~\cite{DBLP:conf/cscw/ResnickISBR94} identifies a set of similar items for each of the items that the user has purchased, and then recommends top-$N$ items based on those similar items.
There are plenty of ways to measure user/item similarity, e.g., Pearson correlation, cosine similarity, and so forth.

The second category is model-based collaborative filtering, in which the latent factor models have achieved the state-of-the-art performance. Cremonesi et.al. proposed a simple model-based algorithm \textsf{PureSVD}~\cite{DBLP:conf/recsys/CremonesiKT10}, where users' features and items' features are represented by the principle singular vectors of the user-item matrix. Koren proposed the well-known \textsf{SVD++} model~\cite{DBLP:conf/recsys/CremonesiKT10}. Wu applied Regularized Matrix Factorization (\textsf{RMF}), Maximum Margin Matrix Factorization (\textsf{MMMF}), and Nonnegative Matrix Factorization (\textsf{NMF}) to recommender systems~\cite{wu2007collaborative}. Weighted Regularized Matrix Factorization (\textsf{WRMF}) was introduced by Hu et al.~\cite{DBLP:conf/icdm/HuKV08}. The key idea of these methods is to factorize the user-item matrix to represent the users’ preferences and items’ characteristics in a common latent space, and then estimate the user-item matrix by the dot product of user factors and item factors. All these methods assume that only a few variables impact users’ preference and items’ features, which means the low-rank structure of user-item matrix.

Another model-based method, \textsf{SLIM}, proposed by Ning et. al.~\cite{DBLP:conf/icdm/NingK11}, predicts the user-item matrix by multiplying the observed user-item matrix by the aggregation coefficient matrix. \textsf{SLIM} estimates the coefficient matrix by learning from the observed user-item matrix with a simultaneous regression model. Specifically, it introduces sparsity with $\ell_1$-norm regularizer into the regularized optimization and formed an elastic net problem to benefit from the smoothness of $\ell_2$-norm and the sparsity of $\ell_1$-norm. Later, plenty of research has been done based on \textsf{SLIM}. \textsf{SSLIM}~\cite{DBLP:conf/recsys/NingK12} integrates the side information. \textsf{LorSLIM}~\cite{DBLP:conf/icdm/ChengYY14} involves the nuclear-norm to induce the low-rank property of \textsf{SLIM}. \textsf{HOSLIM}~\cite{DBLP:conf/pakdd/ChristakopoulouK14} uses the potential higher-order information to generate better recommendation.

The last category is hybrid methods. Hybrid method is used to combine the virtue of different recommender algorithms to generate better performance. A hybrid method was used to deal with the cold-start scenarios by mapping entities (user/item attributes) to latent features of a matrix factorization model~\cite{DBLP:conf/icdm/GantnerDFRS10}. 

%% file: sim.tex
\section{Initializing Item Similarity} \label{sec:network}

This section present the method for measuring item similarity.

\subsection{Preliminaries}
\begin{definition} [Information Network]
An information network is defined as a directed graph $G=(V,E)$ with an object type mapping function $\sigma:V\rightarrow \mathcal{A}$ and a link type mapping function $\varphi:E\rightarrow \mathcal{R}$, where each object $v\in V$ belongs to one particular object type $\sigma(v)\in \mathcal{A}$, each link $e\in E$ belongs to a particular relation $\varphi(e)\in \mathcal{R}$, and if two links belong to the same relation type, the two links share the same starting object type as well as the ending object type. Note that, if a relation exists from type $A$ to type $B$, denoted as $ARB$, the inverse relation $R^{-1}$ holds naturally for $BR^{-1}A$. $R$ and its inverse $R^{-1}$ are usually not equal, unless the two types are the same and $R$ is symmetric. When the types of objects $|\mathcal{A}| > 1$ or the types of relations $|\mathcal{R}| > 1$, the network is called \textbf{heterogeneous} information network; otherwise, it is a \textbf{homogeneous} information network.
\end{definition}

\begin{definition} [Network Schema]
The network schema, denoted as $T_G=(\mathcal{A},\mathcal{R})$, is a meta template for a heterogeneous network $G=(V,E)$ with the object type mapping $\sigma:V\rightarrow\mathcal{A}$ and the link mapping $\varsigma:E\rightarrow\mathcal{R}$, which is a directed graph defined over object types $\mathcal{A}$, with edges as
relations from $\mathcal{R}$.
\end{definition}

\begin{definition} [Meta-path]
A meta-path $\mathcal{P}$ is a path defined on the graph of network schema $T_G=(\mathcal{A},\mathcal{R})$, and is denoted in the form of $A_1\pathto{R_1}A_2\pathto{R_2}\ldots\pathto{R_l}A_{l+1}$. For simplicity, the meta-path can be denoted by the type names if there exist no multiple relations between the same pair of types: $\mathcal{P}=(A_1 A_2 \ldots A_{l+1})$. A path $p=(a_1 a_2\ldots a_{l+1})$ between $a_1$ and $a_{l+1}$ is said to follow the meta-path $\mathcal{P}$, if $\forall i, a_i\in A_i$ and each link $e_i=\pair{a_i a_{i+1}}$ belongs to each relation $R_i$ in $\mathcal{P}$. These paths are called path \emph{instances} of $\mathcal{P}$, denoted as $p\in\mathcal{P}$.
\end{definition}

In the following, meta-path is confined to symmetric, namely \emph{round trip meta-paths} in the form of $\mathcal{P}=(\mathcal{P}_l\mathcal{P}_l^{-1})$.

\begin{definition} [Commuting Matrix] \label{def:matrix}
Given a network $G=(V,E)$ and its network schema $T_G$, a commuting matrix $M$ for a meta-path $\mathcal{P}=(A_1A_2\ldots A_l)$ is defined as $M = W_{A_1A_2}W_{A_2A_3}W_{A_{l-1}A_l}$  , where $W_{A_iA_j}$ is the weight matrix between type $A_i$ and type $A_j$. 
\end{definition}

\begin{definition} [\pathsim] \label{def:pathsim}
Given a symmetric meta-path $\mathcal{P}$, \pathsim between two objects $v_i$ and $v_j$ of the same type is:
  \begin{equation}\label{eq:pathsim}
    s_{ij}=\frac{2M_{ij}}{M_{ii}+M_{jj}},
  \end{equation}
  where $M_{ij}$ represents the $i^{th}$ row and $j^{th}$ column element of matrix $M$.
\end{definition}

\subsection{Measuring Item Similarity}
To measure item similarity through \pathsim, we first define the meta-path in the form of $\mathcal{P}_n=(A(BA)^n)$ (the mined frequent patterns~\cite{DBLP:conf/icdm/ChenZLW15}). For instance, $n=1$ corresponds to $\mathcal{P}_1=(ABA)$ and $n=2$ corresponds to $\mathcal{P}_2=(ABABA)$. It is easy to verify the symmetry of $\mathcal{P}_n$ and thus \pathsim can be applied. The associated commuting matrix for $\mathcal{P}_n$ is $M=(W_{AB}W_{BA})^n$ and consequently the similarity between item $i$ and item $j$ can be computed by Equation~(\ref{eq:pathsim}).

Suppose we define $N$ meta-paths $\mathcal{P}_1,\mathcal{P}_2,\ldots,\mathcal{P}_N$, with the corresponding similarities $s_1,s_2,\ldots,s_N$, the overall similarity should be measured as the weighted aggregation, e.g. $s=\sum_{n=1}^N \alpha_n s_n$, where $\sum_{n=1}^N \alpha_n=1$. As is suggested in \cite{DBLP:journals/pvldb/SunHYYW11} that the meta-path with relatively short length is good enough to measure similarity, and a long meta-path may even reduce the quality, we set smaller weights for longer meta-paths. We naturally set the weights as $ \alpha_n=\frac{2^{N-n}}{2^N-1}$. We further denote $\vec{S}^p$ for the matrix of \pathsim, where $s^p_{ij}$ represents the element of $\vec{S}^p$ in the $i^{th}$ row and $j^{th}$ column. 

%% file: learn.tex
\section{Optimizing Profile Similarity} \label{sec:learn}
We prompt to measure the item similarity based on the profiles. Prior to the discussion, we first list the notations used in this section in Table~\ref{tab:notation}. Note that vectors and matrices are made \textbf{bold}.

\begin{table}
  \centering
  \caption{Table of Notations}\label{tab:notation}
  \begin{tabular}{|c|c|}
  \hline
  $N_u$ & Number of users \\
  \hline
  $N_v$ & Number of items \\
  \hline
  $N_w$ & Number of terms \\
  \hline
  $\lambda$ & $\ell_2$ norm weight \\
  \hline
  $\vec{S}^f,s_{ij}^f$ & similarity derived from item profiles \\
  \hline
  $\vec{S}^p,s_{ij}^p$ & similarity derived form pathsim \\
  \hline
  $\vec{W}^l,\vec{w}_i^l,w_{ik}^l$ & local term weights \\
  \hline
  $\vec{w}^g,w_{ik}^g$ & global term weights \\
  \hline
  $\vec{w},w_k$ & the weights to learn, where $w_k=(w_k^g)^2$\\
  \hline
  $\vec{P},\vec{p}_k$ & the normalized $tf\times idf$ weights \\
  \hline
\end{tabular}
\end{table}

Each profile contains rich text to describe the feature. Thus more effective content analysis methods and text similarity measures are crucial for the recommendation. Most designed recommender systems involving text similarity measure applied cosine similarity of two bags of words, where each word is weighted by $tf\times idf$~\cite{DBLP:conf/recsys/BarjastehFMER15,DBLP:conf/recsys/SaveskiM14}. Nevertheless, it is possible to go beyond the definition of $tf\times idf$, where $tf$ represents the local term weights and $idf$ the global term weights. While $tf$ could be derived offline with various methods, $idf$ requires further optimization as suggested by~\cite{DBLP:conf/www/GuZHS16}. Thus the global term could be optimized with the guidance of \pathsim and the similarity derived from profiles could be calculated by the following Equation:
\begin{equation}\label{eq:profile-sim}
  s_{ij}^f=\frac{\vec{d}_i\cdot \vec{d}_j}{\norm{\vec{d}_i}_2\norm{\vec{d}_j}_2},
\end{equation}
where $\norm{\cdot}_2$ is the $\ell_2$ norm of a vector, and $\vec{d}_i$ represents the term vector, each dimension represents a term, and the value in each dimension represents the weight of the term. $\vec{d}_i$ could be decomposed as $\vec{w}^{l}_i\circ\vec{w}^{g}$, where $\vec{w}^g_i$ denotes for the local weights for item $i$. $\vec{w}^g$ denotes for the global term weights, which is initially set with the original Inverted Document Frequency, and optimized gradually. $\circ$ is a binary operation, conducting the element-wise product of two vectors, thus the result is also a vector. By letting $w_k=(w^g_k)^2$, we could further formalize Equation~\ref{eq:profile-sim} as follows:

\begin{displaymath}
  s^f_{ij}=\frac{\sum_{k=1}^t w_{ik}^l w_{jk}^l w_k}{\left[\sum_{k=1}^t (w_{ik}^l)^2 w_k\right]^{\frac{1}{2}} \left[\sum_{k=1}^t (w_{jk}^l)^2 w_k\right]^{\frac{1}{2}}},
\end{displaymath}
and the partial derivative could be derived as:
\begin{displaymath}
\begin{split}
   \derivation{s^f_{ij}}{w_k} &=\frac{1}{\norm{\vec{d}_i}_2^2\norm{\vec{d}_j}_2^2} \left\{w_{ik}^l w_{jk}^l \norm{\vec{d}_i}_2\norm{\vec{d}_j}_2-\left[\frac{\norm{\vec{d}_j}_2}{2\norm{\vec{d}_i}_2} (w_{ik}^l)^2+\frac{\norm{\vec{d}_i}_2}{2\norm{\vec{d}_j}_2} (w_{jk}^l)^2\right]\vec{d}_i\cdot\vec{d}_j \right\} \\
   &=\frac{w_{ik}^l w_{jk}^l}{\norm{\vec{d}_i}_2 \norm{\vec{d}_j}_2}-\frac{s^f_{ij}}{2}\left[\frac{(w_{ik}^l)^2}{\norm{\vec{d}_i}_2^2}+\frac{(w_{jk}^l)^2}{\norm{\vec{d}_j}_2^2} \right].
\end{split}
\end{displaymath}
%

To optimize the global term weights, we should define the loss function to measure the difference between $s^p_{ij}$ and $s^f_{ij}$. We develop the squared loss function and the associated optimization methods. 

\subsection{Squared Error Loss Function}

Due to the sparsity of the user-item information network, we could also expect the sparsity of similarities measured by \pathsim. If item $i$ can not reach item $j$ through the bi-type information network, according to Equation~\ref{eq:pathsim}, $s_{ij}=0$. 

In this section, the loss function is defined as the squared error, given by
\begin{displaymath}
  \mathcal{L}=\sum_{i=1}^{N_v}\sum_{j=1}^{N_v} (s^f_{ij}-s^p_{ij})^2=\norm{\vec{S}^f-\vec{S}^p}_F^2,
\end{displaymath}
based on which, we could minimize the following objective function to optimize the global term frequency:
\begin{equation}\label{eq:matrix-obj}
\begin{split}
   \min_{\vec{w}} J &=\frac{1}{2}\norm{\vec{S}^f-\vec{S}^p}_F^2+\frac{\lambda}{2}\norm{\vec{w}}_2^2 \\
   &s.t.~ \mat{w}\geq 0
\end{split}
\end{equation}
where $\norm{\cdot}_F$ is the Frobenius norm, which is actually the squared sum of all elements of the matrix. $\vec{w}$ stands for the vector of $w_k$, and we penalize $\ell_2$ norm on the global term weights $\vec{w}$ to avoid over fitting and sparsity result. $\vec{S}^p$ is denoted for \pathsim matrix, whereas $\vec{S}^f$ for the profile similarity matrix. 
We reformulate the problem into the following element-wise form, to facilitate the deduction of partial derivative over $w_k$, e.g., $\derivation{J}{w_k}$.

\begin{displaymath}
\begin{split}
   \min_{w_k} J &=\frac{1}{2}\sum_{i=1}^{N_v}\sum_{j=1}^{N_v} (s^f_{ij}-s^p_{ij})^2+\frac{\lambda}{2}\sum_{i=1}^{N_w} w_k^2 \\
   & ~w_k\geq 0,k=1,\ldots,N_w
\end{split}
\end{displaymath}

\myparagraph{Solution}
The partial derivative is given in Equation~\ref{eq:derivation}.
\begin{equation}\label{eq:derivation}
  \derivation{J}{w_k}=\sum_{i=1}^{N_v}\sum_{j=1}^{N_v} (s^f_{ij}-s^p_{ij}) \left\{ \frac{w_{ik}^l w_{jk}^l}{\norm{\vec{d}_i}_2 \norm{\vec{d}_j}_2}-\frac{\tilde{s}_{ij}}{2}\left[\frac{(w_{ik}^l)^2}{\norm{\vec{d}_i}_2^2}+\frac{(w_{jk}^l)^2}{\norm{\vec{d}_j}_2^2} \right] \right\} + \lambda w_k.
\end{equation}

We further define $q_{ij}=s^f_{ij}-s^p_{ij}, p_{ik}=\frac{w_{ij}^l}{\norm{\vec{d}_i}_2}$ and $r_{ij}=(s^f_{ij}-s^p_{ij})s^f_{ij}$. Thus we have:
\begin{displaymath}
 \sum_{i=1}^{N_v}\sum_{j=1}^{N_v}(s^f_{ij}-s^p_{ij})\frac{w_{ik}^l w_{jk}^l}{\norm{\vec{d}_i}_2 \norm{\vec{d}_j}_2}= \sum_{i=1}^{N_v}\sum_{j=1}^{N_v}q_{ij}p_{ik}p_{jk}=\vec{p}_k^T\vec{Q}\vec{p}_k
\end{displaymath}
\begin{displaymath}
\begin{split}
&\sum_{i=1}^{N_v}\sum_{j=1}^{N_v}(s^f_{ij}-s^p_{ij})\frac{s^f_{ij}}{2}\left[ \frac{(w_{ik}^l)^2}{\norm{\vec{d}_i}_2^2}+\frac{(w_{jk}^l)^2}{\norm{\vec{d}_j}_2^2} \right]=\sum_{i=1}^{N_v}\sum_{j=1}^{N_v}\frac{1}{2}r_{ij}(p_{ik}^2+p_{jk}^2) \\
&=\sum_{i=1}^{N_v}\sum_{j=1}^{N_v}r_{ij}p_{ik}^2=\vec{p}_k^T\vec{R}\vec{p}_k.
\end{split}
\end{displaymath}
where $\vec{p}_k$ is a vector of $p_{ik}$, $\vec{Q}$ is $N_v\times N_v$ matrix of $q_{ij}$ and $\vec{R}$ is a diagonal matrix with the $i$-th element of principal diagonal equals $\sum_j r_{ij}$. By defining $\vec{L}=\vec{Q}-\vec{R}$, we find the following close form of derivative:
\begin{displaymath}
  \derivation{J}{w_k}=\vec{p}_k^T\vec{L}\vec{p}_k+\lambda w_k.
\end{displaymath}
It could be further represented into the matrix form:
\begin{equation}\label{eq:close}
  \derivation{J}{\vec{w}}=\diag{\vec{P}^T\vec{LP}}+\lambda \vec{w},
\end{equation}
where $\diag{\cdot}$ extracts the principal diagonal and form as a vector.

Following the common practices for top-$N$ recommendation~\cite{DBLP:conf/kdd/KabburNK13}, the loss function is computed over all entries of $\vec{S}$. The summation above contains $n\times n$ terms, namely all pairwise items in the dataset. To ensure good performance while achieve reasonable training time, the algorithm is paralleled by CUDA.

%% file: exp.tex
\section{Experimental Evaluation} \label{sec:exp}

To evaluate our proposed method, extensive experiments have been conducted. However, due to space limitation, we only present part of the results.

\subsection{Experiment Setup}
The results reported in this section is based on the NIPS dataset\myfootnote{http://www.cs.nyu.edu/\~{}roweis/data.html}.
It contains paper-author and paper-word matrices extracted from co-author network at the NIPS conference over 13 volumes. We regard authors as users, papers as items and the contents of papers as the profile of items. Thus the data has 2037 users (authors) and 1740 items (papers), where 13649 words have been extracted from the corpus of item profiles. The content of the papers is preprocessed such that all words are converted to lower case and stemmed and stop-words are removed. One may note that NIPS dataset is very sparse, that is, some author may publish only one or two papers, which shows the importance of properly leveraging side information for recommendation.

We applied 5-time Leave-One-Out cross validation (LOOCV) to evaluate our proposed method. In each run, each of the dataset is split into a training set and a testing set by randomly selecting one of the non-zero entries of each user and placing it into the testing set. The training set is used to train a model, then for each user a size-$N$ ranked list of recommended items is generated by the model. We varies $N$ as 5,10,15,20 to compare the result difference. Our method has two parameters, $n_p$ and $\lambda$. $n_p$ measures the length of meta-path and $\lambda$ measures the degree of regularization.

The recommendation quality is measured using Hit Rate (HR) and Average Reciprocal Hit Rank (ARHR)~\cite{DBLP:journals/tois/DeshpandeK04}. HR is defined as
\begin{displaymath}
  HR=\frac{\#hits}{\#users},
\end{displaymath}
where $\#users$ is the total number of users and $\#hits$ is the number of users whose item in the testing set is recommended (i.e., hit) in the size-$N$ recommendation list. A second measure
for evaluation is ARHR, which is defined as
\begin{displaymath}
  ARHR=\frac{1}{\#users}\sum_{i=1}^{\#hits}\frac{1}{p_i},
\end{displaymath}
where if an item of a user is hit, $p$ is the position of the item in the ranked recommendation list. ARHR is a weighted version of HR and it measures how strongly an item is recommended, in which the weight is the reciprocal of the hit position in the recommendation list.

We implement our algorithm in C++. As our method involves optimizing the global weights over the whole vocabulary of item profiles, to expedite the training efficiency, the training process is paralleled in GPU and implemented by CUDA\myfootnote{http://www.nvidia.cn/object/cuda-cn.html}. All experiments are done on a machine with 4-core Intel i7-4790 processor at 3.60GHz and Nvidia GeForce GTX TITAN X graphics card.

\subsection{Effect of Initial Value}
\begin{figure}
  \centering
  \subfigure[HR]{
  \label{fig:initial-hr}
  \includegraphics[width=0.475\textwidth]{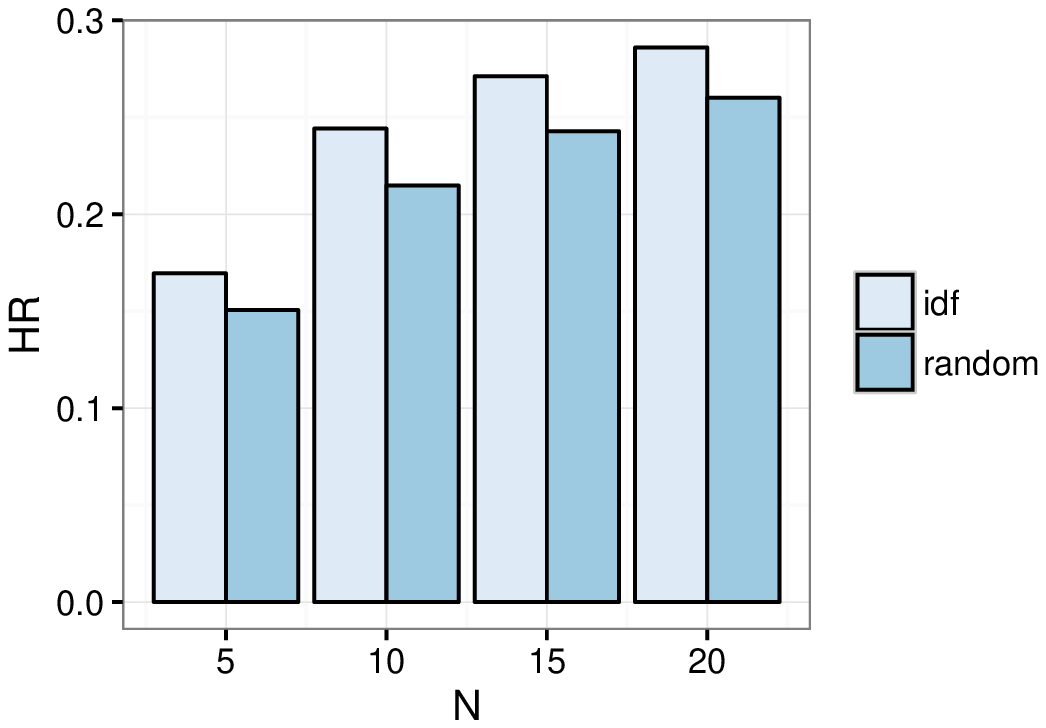}
  }
  \subfigure[ARHR]{
  \label{fig:initial-arhr}
  \includegraphics[width=0.475\textwidth]{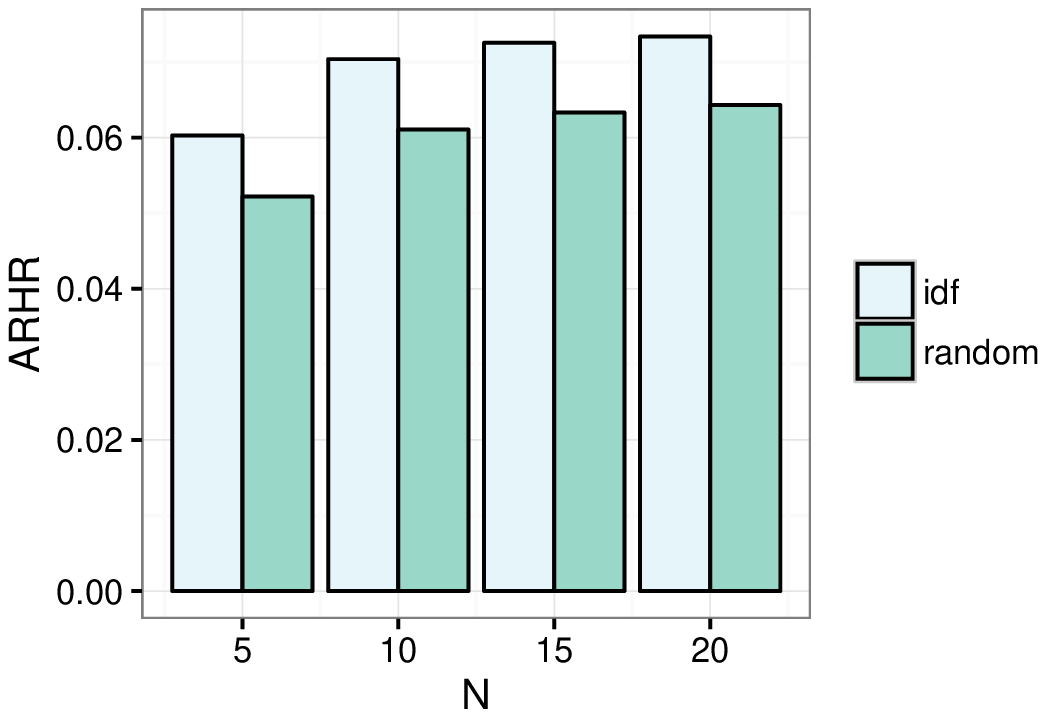}
  }
  \caption{Effect of Initial Value}\label{fig:initial}
\end{figure}

We first evaluate the influence of initial value we set for global weights on the performance. We compare two settings, \textsf{random} and \textsf{idf}. The initial value is randomly set in the first setting while it is set as the value of inverted document frequency in the latter one. Here $\lambda$ is set as 0.01 and $n_p$ as 1. The result is reported in Figure~\ref{fig:initial}, which shows the superiority of \textsf{idf} over \textsf{random}. The result demonstrates the usefulness of side information in this dataset and we set the initial value of global term as \textsf{idf} thereafter.

\subsection{Effect of Parameters}\label{sec:parameter}
\begin{figure}[t]
  \centering
  \subfigure[HR]{
  \label{fig:lambda-hr}
  \includegraphics[width=0.475\textwidth]{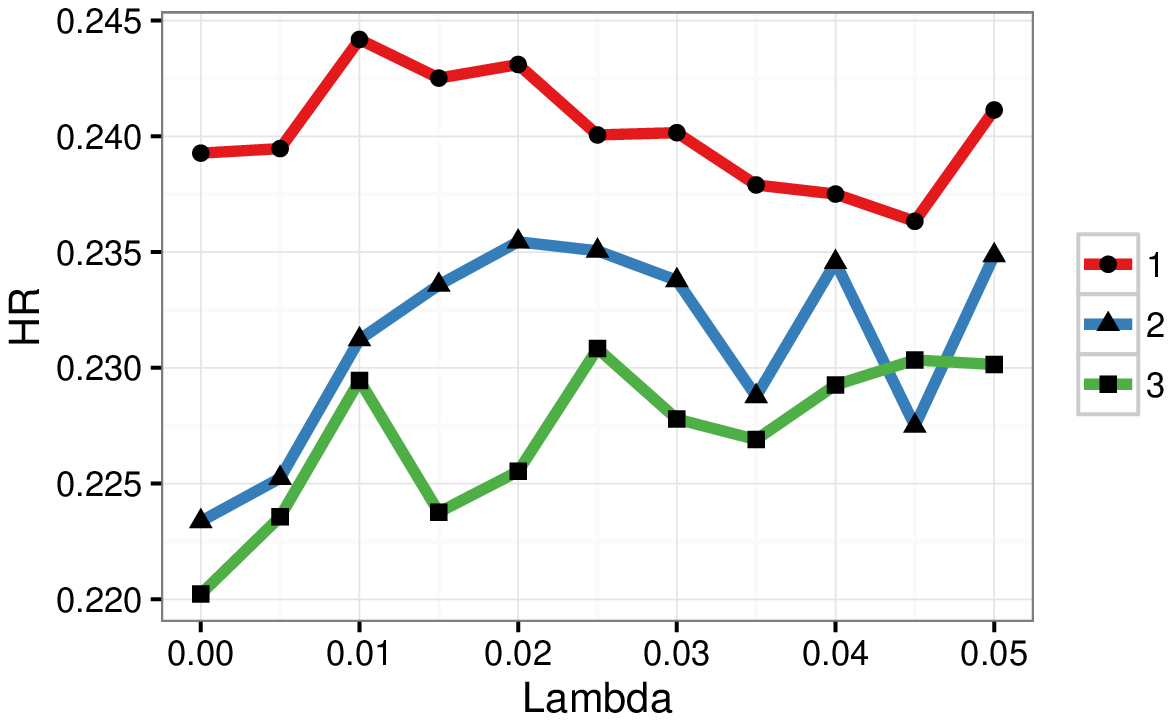}
  }
  \subfigure[ARHR]{
  \label{fig:lambda-arhr}
  \includegraphics[width=0.475\textwidth]{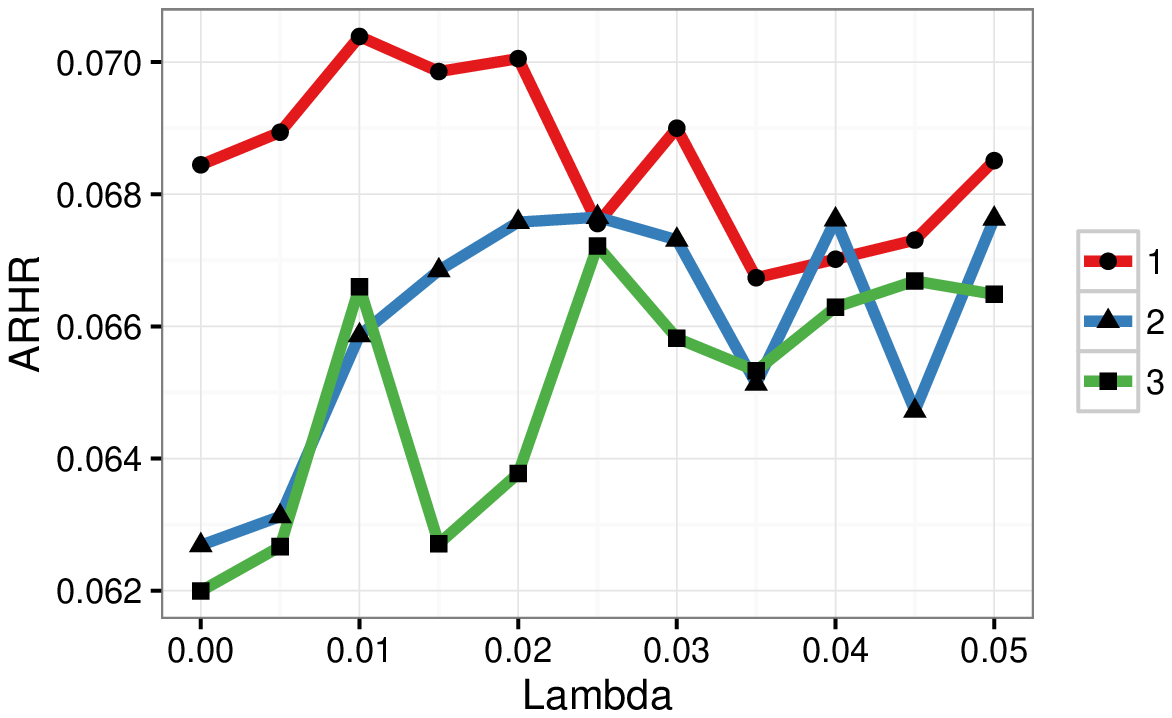}
  }
  \caption{Effect of Parameter $\lambda$ and $n_p$}\label{fig:parameter}
\end{figure}

In this set of experiments, the validation is conducted to select the most suitable parameters. $\lambda$ is varied from $0$ to $0.05$ and stepped $0.005$ and $n_p$ is set as 1,2,3. We draw the lines in Figure~\ref{fig:parameter}. Three lines are drawn to distinguish $n_p=1$ (red line),$n_p=2$ (blue line) and $n_p=3$ (green line) respectively. The result shows that $n_p$ should be set 1 to achieve better performance. This result is consistent with~\cite{DBLP:journals/pvldb/SunHYYW11}, which suggests shorter length of meta-path is good enough to measure similarity.

As Figure~\ref{fig:parameter} depicts the performance along with $\lambda$, we find the best value as $0.01$ for $n_p=1$, $0.02$ for $n_p=1$ and $0.025$ for $n_p=3$. It has also been shown that the method performs more robustly when $n_p=1$ while it varies dramatically with $\lambda$ when $n_p>1$. Based on the observation, we finally pick $n_p$ and $\lambda$ as 1 and 0.01 for the rest of the experiments.


\subsection{Recommendation for Different Top-$N$}
\begin{figure}
  \centering
  \subfigure[HR]{
  \label{fig:topn-hr}
  \includegraphics[width=0.475\textwidth]{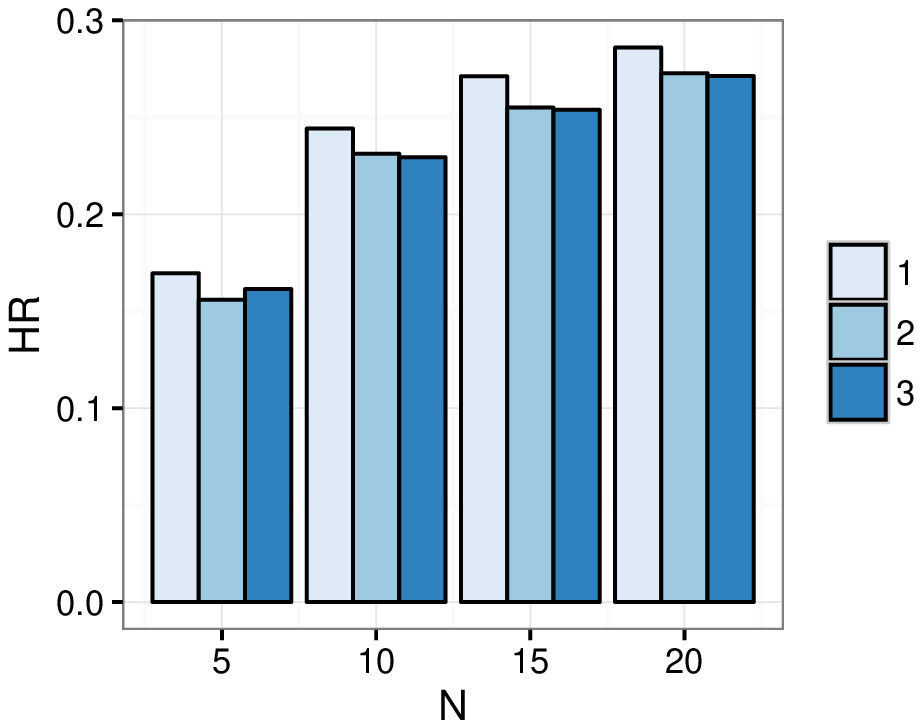}
  }
  \subfigure[ARHR]{
  \label{fig:topn-arhr}
  \includegraphics[width=0.475\textwidth]{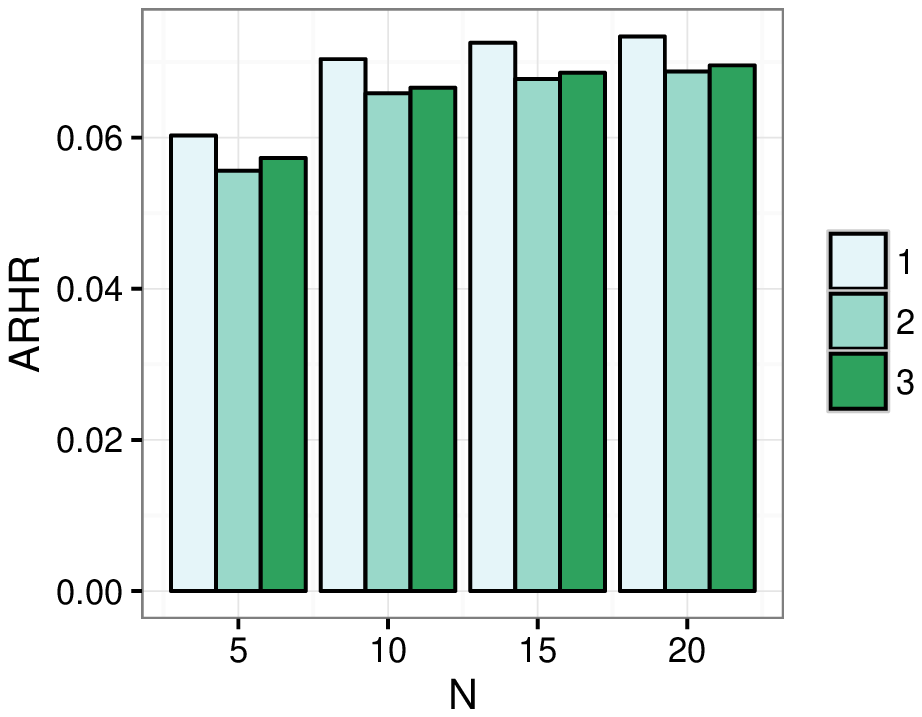}
  }
  \caption{Performance of Proposed Method}\label{fig:performance}
\end{figure}

By setting the global term as the inverted document frequency, and letting $\lambda=0.01$, we evaluate the top-$N$ recommendation performance, the result of which is illustrated in Figure~\ref{fig:performance}. Obviously, with the increase of $N$, the performance improves. We also compare the different setting of $n_p$, which further demonstrates $n_p=1$ could be a better choice. We also found in this set of experiments that when $N$ increase from 5 to 10, the performance shows relatively higher improvement. 

\subsection{Comparison of Algorithms}
\begin{figure}
  \centering
  \subfigure[HR]{
  \label{fig:compare-hr}
  \includegraphics[width=0.475\textwidth]{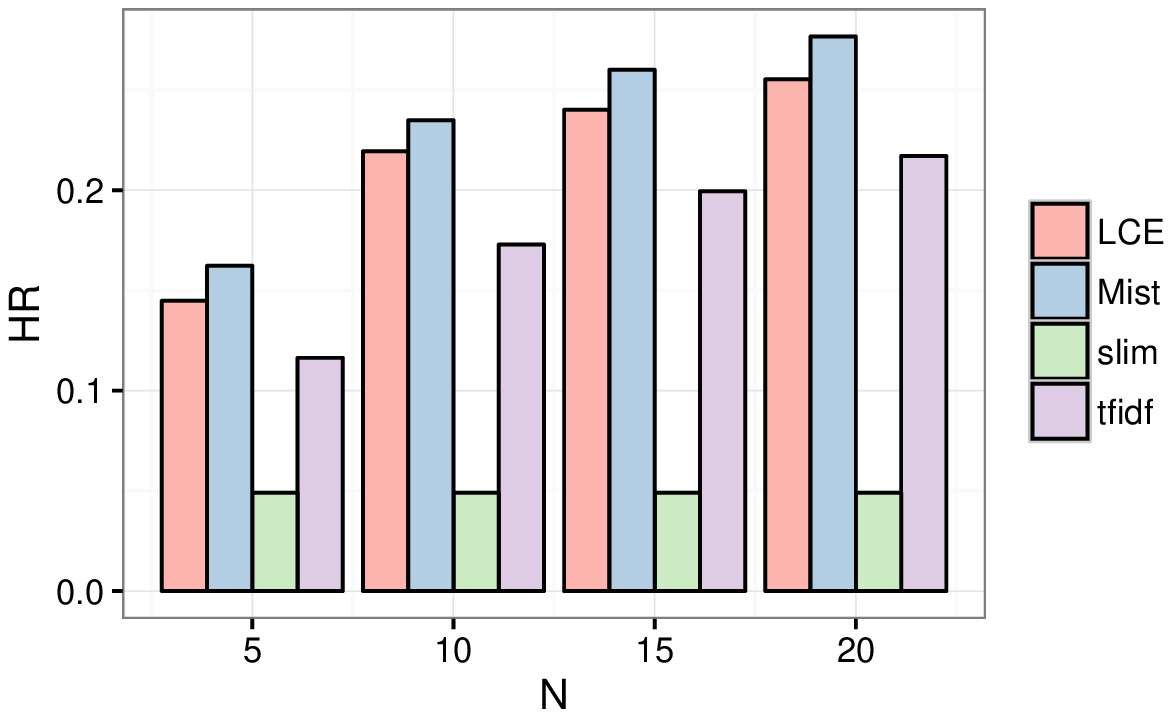}
  }
  \subfigure[ARHR]{
  \label{fig:compare-arhr}
  \includegraphics[width=0.475\textwidth]{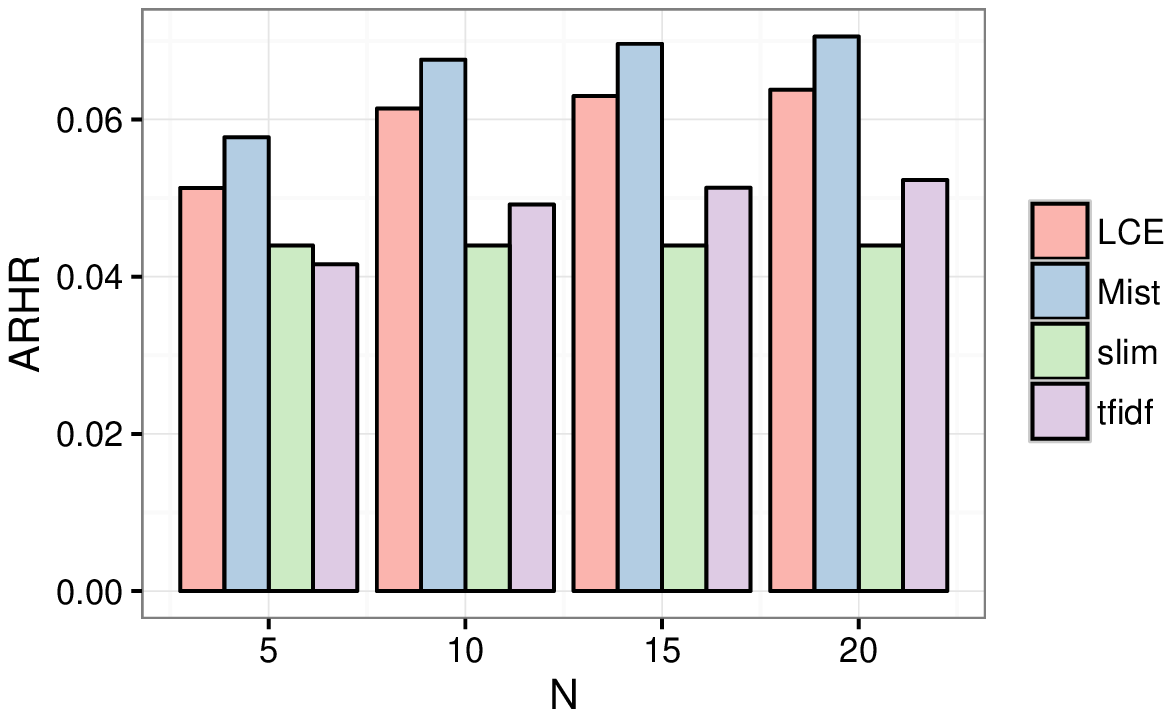}
  }
  \caption{Algorithm Comparison}\label{fig:comapre}
\end{figure}

We finally compares our method with other algorithms in this set of experiments. As top-$N$ recommendation methods have been extensively studied, we compare only with some state-of-the-art methods, e.g. \textsf{Slim}~\cite{DBLP:conf/icdm/NingK11} and \textsf{LCE}~\cite{DBLP:conf/recsys/SaveskiM14}. We also incorporate the pure \textsf{tfidf} method to calculate the item similarity for recommendation. To distinguish, we name our proposed method as \textsf{Mist} (\underline{M}eta path based \underline{i}tem \underline{s}imilarity to learn global \underline{t}erm weights). We depict the result in Figure~\ref{fig:comapre}, where all the compared algorithms were optimized to the best settings.

Figure~\ref{fig:compare-hr} shows the recommendation of \textsf{Mist} is consistently better than other three methods. Note that \textsf{Slim} has the worst performance, this could be attributed to the sparsity of dataset. \textsf{LCE} also took advantage of item profiles, thus it has achieved good performance. It is also worth noting that the pure \textsf{tfidf} shows relatively acceptable results and \textsf{Mist} could be regarded as the collaborative optimized \textsf{tfidf}.

When it comes to ARHR, showing in Figure~\ref{fig:compare-arhr}, \textsf{Mist} also behaves the best, which was followed by \textsf{LCE}, \textsf{tfidf} and \textsf{Slim}. In conclusion, the learned global term weights can well capture both structural and textual information.

%% file: recsys.bbl
\begin{thebibliography}{10}
\providecommand{\url}[1]{\texttt{#1}}
\providecommand{\urlprefix}{URL }

\bibitem{balabanovic1997fab}
Balabanovi{\'c}, M., Shoham, Y.: Fab: content-based, collaborative
  recommendation. Communications of the ACM  40(3),  66--72 (1997)

\bibitem{DBLP:conf/recsys/BarjastehFMER15}
Barjasteh, I., Forsati, R., Masrour, F., Esfahanian, A., Radha, H.: Cold-start
  item and user recommendation with decoupled completion and transduction. In:
  Proceedings of the 9th {ACM} Conference on Recommender Systems, RecSys 2015,
  Vienna, Austria, September 16-20, 2015. pp. 91--98 (2015)

\bibitem{DBLP:conf/icdm/ChenZLW15}
Chen, Y., Zhao, X., Lin, X., Wang, Y.: Towards frequent subgraph mining on
  single large uncertain graphs. In: 2015 {IEEE} International Conference on
  Data Mining, {ICDM} 2015, Atlantic City, NJ, USA, November 14-17, 2015. pp.
  41--50 (2015)

\bibitem{DBLP:conf/icdm/ChengYY14}
Cheng, Y., Yin, L., Yu, Y.: {LorSLIM}: Low rank sparse linear methods for top-n
  recommendations. In: 2014 {IEEE} International Conference on Data Mining,
  {ICDM} 2014, Shenzhen, China, December 14-17, 2014. pp. 90--99 (2014)

\bibitem{DBLP:conf/pakdd/ChristakopoulouK14}
Christakopoulou, E., Karypis, G.: {HOSLIM:} higher-order sparse linear method
  for top-n recommender systems. In: Advances in Knowledge Discovery and Data
  Mining - 18th Pacific-Asia Conference, {PAKDD} 2014, Tainan, Taiwan, May
  13-16, 2014. Proceedings, Part {II}. pp. 38--49 (2014)

\bibitem{DBLP:conf/recsys/CremonesiKT10}
Cremonesi, P., Koren, Y., Turrin, R.: Performance of recommender algorithms on
  top-n recommendation tasks. In: Proceedings of the 2010 {ACM} Conference on
  Recommender Systems, RecSys 2010, Barcelona, Spain, September 26-30, 2010.
  pp. 39--46 (2010)

\bibitem{DBLP:journals/tois/DeshpandeK04}
Deshpande, M., Karypis, G.: Item-based top-\emph{N} recommendation algorithms.
  {ACM} Trans. Inf. Syst.  22(1),  143--177 (2004)

\bibitem{DBLP:conf/icdm/GantnerDFRS10}
Gantner, Z., Drumond, L., Freudenthaler, C., Rendle, S., Schmidt{-}Thieme, L.:
  Learning attribute-to-feature mappings for cold-start recommendations. In:
  {ICDM} 2010, The 10th {IEEE} International Conference on Data Mining, Sydney,
  Australia, 14-17 December 2010. pp. 176--185 (2010)

\bibitem{DBLP:conf/www/GuZHS16}
Gu, Y., Zhao, B., Hardtke, D., Sun, Y.: Learning global term weights for
  content-based recommender systems. In: Proceedings of the 25th International
  Conference on World Wide Web, {WWW} 2016, Montreal, Canada, April 11 - 15,
  2016. pp. 391--400 (2016)

\bibitem{DBLP:conf/icdm/HuKV08}
Hu, Y., Koren, Y., Volinsky, C.: Collaborative filtering for implicit feedback
  datasets. In: Proceedings of the 8th {IEEE} International Conference on Data
  Mining {(ICDM} 2008), December 15-19, 2008, Pisa, Italy. pp. 263--272 (2008)

\bibitem{DBLP:conf/kdd/KabburNK13}
Kabbur, S., Ning, X., Karypis, G.: {FISM:} factored item similarity models for
  top-n recommender systems. In: The 19th {ACM} {SIGKDD} International
  Conference on Knowledge Discovery and Data Mining, {KDD} 2013, Chicago, IL,
  USA, August 11-14, 2013. pp. 659--667 (2013)

\bibitem{DBLP:conf/cikm/Karypis01}
Karypis, G.: Evaluation of item-based top-n recommendation algorithms. In:
  Proceedings of the 2001 {ACM} {CIKM} International Conference on Information
  and Knowledge Management, Atlanta, Georgia, USA, November 5-10, 2001. pp.
  247--254 (2001)

\bibitem{DBLP:conf/icdm/NingK11}
Ning, X., Karypis, G.: {SLIM:} sparse linear methods for top-n recommender
  systems. In: 11th {IEEE} International Conference on Data Mining, {ICDM}
  2011, Vancouver, BC, Canada, December 11-14, 2011. pp. 497--506 (2011)

\bibitem{DBLP:conf/recsys/NingK12}
Ning, X., Karypis, G.: Sparse linear methods with side information for top-n
  recommendations. In: Sixth {ACM} Conference on Recommender Systems, RecSys
  '12, Dublin, Ireland, September 9-13, 2012. pp. 155--162 (2012)

\bibitem{papagelis2005qualitative}
Papagelis, M., Plexousakis, D.: Qualitative analysis of user-based and
  item-based prediction algorithms for recommendation agents. Engineering
  Applications of Artificial Intelligence  18(7),  781--789 (2005)

\bibitem{DBLP:conf/cscw/ResnickISBR94}
Resnick, P., Iacovou, N., Suchak, M., Bergstrom, P., Riedl, J.: Grouplens: An
  open architecture for collaborative filtering of netnews. In: {CSCW} '94,
  Proceedings of the Conference on Computer Supported Cooperative Work, Chapel
  Hill, NC, USA, October 22-26, 1994. pp. 175--186 (1994)

\bibitem{DBLP:conf/recsys/SaveskiM14}
Saveski, M., Mantrach, A.: Item cold-start recommendations: learning local
  collective embeddings. In: Eighth {ACM} Conference on Recommender Systems,
  RecSys '14, Foster City, Silicon Valley, CA, {USA} - October 06 - 10, 2014.
  pp. 89--96 (2014)

\bibitem{DBLP:journals/pvldb/SunHYYW11}
Sun, Y., Han, J., Yan, X., Yu, P.S., Wu, T.: {PathSim}: Meta path-based top-k
  similarity search in heterogeneous information networks. {PVLDB}  4(11),
  992--1003 (2011)

\bibitem{DBLP:conf/mm/WangLWZ15}
Wang, Y., Lin, X., Wu, L., Zhang, W.: Effective multi-query expansions: Robust
  landmark retrieval. In: Proceedings of the 23rd Annual {ACM} Conference on
  Multimedia Conference, {MM} '15, Brisbane, Australia, October 26 - 30, 2015.
  pp. 79--88 (2015)

\bibitem{DBLP:conf/mm/WangLWZZ14}
Wang, Y., Lin, X., Wu, L., Zhang, W., Zhang, Q.: Exploiting correlation
  consensus: Towards subspace clustering for multi-modal data. In: Proceedings
  of the {ACM} International Conference on Multimedia, {MM} '14, Orlando, FL,
  USA, November 03 - 07, 2014. pp. 981--984 (2014)

\bibitem{DBLP:conf/sigir/WangLWZZ15}
Wang, Y., Lin, X., Wu, L., Zhang, W., Zhang, Q.: {LBMCH:} learning bridging
  mapping for cross-modal hashing. In: Proceedings of the 38th International
  {ACM} {SIGIR} Conference on Research and Development in Information
  Retrieval, Santiago, Chile, August 9-13, 2015. pp. 999--1002 (2015)

\bibitem{DBLP:journals/tip/WangLWZZH15}
Wang, Y., Lin, X., Wu, L., Zhang, W., Zhang, Q., Huang, X.: Robust subspace
  clustering for multi-view data by exploiting correlation consensus. {IEEE}
  Trans. Image Processing  24(11),  3939--3949 (2015)

\bibitem{wang2015unsupervised}
Wang, Y., Zhang, W., Wu, L., Lin, X., Zhao, X.: Unsupervised metric fusion over
  multiview data by graph random walk-based cross-view diffusion. IEEE
  Transactions on Neural Networks and Learning Systems  (2015)

\bibitem{wu2007collaborative}
Wu, M.: Collaborative filtering via ensembles of matrix factorizations. In:
  Proceedings of KDD Cup and Workshop. vol. 2007 (2007)

\end{thebibliography}
